\documentclass[aps,prl,preprintnumbers,showpacs,nofootinbib]{revtex4}
\usepackage{graphicx}
\usepackage{dcolumn}
\usepackage{bm}
\usepackage{subfig}
\usepackage{amsmath, amssymb, graphics}
\begin{document}
\newcommand{\mathsym}[1]{{}}
\newcommand{\unicode}{{}}
\newcommand{\newc}{\newcommand}
\newc{\ra}{\rightarrow}
\newc{\lra}{\leftrightarrow}
\newc{\beq}{\begin{equation}}
\newc{\eeq}{\end{equation}}
\newc{\barr}{\begin{eqnarray}}
\newc{\earr}{\end{eqnarray}}
\def \lta {\mathrel{\vcenter
     {\hbox{$<$}\nointerlineskip\hbox{$\sim$}}}}
\def \gta {\mathrel{\vcenter
     {\hbox{$>$}\nointerlineskip\hbox{$\sim$}}}}
\def\vbf{\mbox{\boldmath $\upsilon$}}
\def\barr{\begin{eqnarray}}
\def\earr{\end{eqnarray}}
\def\g{\gamma}
\newcommand{\dphi}{\delta \phi}
\newcommand{\bupsilon}{\mbox{\boldmath \upsilon}}
\newcommand{\at}{\tilde{\alpha}}
\newcommand{\pt}{\tilde{p}}
\newcommand{\Ut}{\tilde{U}}
\newcommand{\rhb}{\bar{\rho}}
\newcommand{\pb}{\bar{p}}
\newcommand{\pbb}{\bar{\rm p}}
\newcommand{\kt}{\tilde{k}}
\newcommand{\kb}{\bar{k}}
\newcommand{\wt}{\tilde{w}}
\title{The impact of going beyond the Maxwell distribution in direct dark matter detection rates}
\author{J.~D.~Vergados$^{(1)}$\thanks{Vergados@uoi.gr}, S.~H.~Hansen$^{(2)}$ and O.~Host$^{(2)}$ }
\affiliation{$^{(1)}${\it  University
of Cyprus, Nicosia, CY-1678, Cyprus}}
\affiliation{$^{(2)}${\it Dark Cosmology Centre, Niels Bohr Institute, University of Copenhagen, Juliane Maries Vej 30, DK-2100 Copenhagen, Denmark}}

\begin{abstract}
We consider direct dark matter detection rates and investigate the
difference between a standard Maxwell-Boltzmann velocity distribution
and a "realistic" distribution like the ones extracted from numerical
$N$-body simulations. Sizable differences are observed when such
results are compared to the standard Maxwell-Boltzmann
distribution. For a light target both the total rate and the annual
modulation are reduced by $\sim25$\%. For a heavy target the total
rate is virtually unchanged, whereas the annual modulation is modified
by up to 50\%, depending on the WIMP mass and detector energy
threshold. We also consider the effect of a possible velocity
anisotropy, and the effect is found to be largest for a light target
For the realistic velocity distribution the anisotropy may reduce the
annual modulation, in contrast to the Maxwell-Boltzmann case.
\end{abstract}
\date{\today}
\maketitle
   
\section{Introduction}

The universe is observed to contain large amounts of dark matter
\cite{spergel,WMAP06}, and its contribution to the total energy density is
estimated to be $\sim 25\%$. This non-baryonic dark matter component,
responsible for the growth of cosmological perturbations through
gravitational instability, has still not been detected directly. Even
though there exists firm indirect evidence from the halos of dark matter in galaxies
and clusters of galaxies it is
essential to detect matter directly.

The possibility of direct detection, however, depends on the nature of
the dark matter constituents. Supersymmetry naturally provides
candidates for these constituents
\cite{goodwit,KVprd,ellrosz,ref2}. In the most favored scenario of
supersymmetry, the lightest supersymmetric particle (LSP) can be
described as a Majorana fermion, a linear combination of the neutral
components of the gauginos and higgsinos.
 
Since the LSPs (or WIMPs) are expected to be very massive
($m_{\text{WIMP}} \gta 30$ GeV) and extremely non-relativistic with
average kinetic energy $ \langle T \rangle \simeq$ 50 keV
$\left(m_{\text{WIMP}}/ 100\, {\rm GeV} \right)$, a WIMP interaction
with a nucleus in an underground detector is not likely to produce
excitation.  As a result, WIMPs can be directly detected mainly via
the recoil of a nucleus ($A,Z$) in elastic scattering. The event rate
for such a process can be computed from the following ingredients:
\begin{enumerate}
\item An effective Lagrangian at the elementary particle (quark)
level obtained in the framework of the prevailing particle theory. For supersymmetry 
this is achieved as described in refs. \cite{ref2,JDV96}, for example.
\item A well defined procedure for transforming the amplitude
obtained using the previous effective Lagrangian from the quark to
the nucleon level, i.e. a quark model for the nucleon. This step
in SUSY models is non-trivial, since the obtained results depend crucially on the
content of the nucleon in quarks other than u and d.
\item Knowledge of the relevant nuclear matrix elements
\cite{Ress,DIVA00}, obtained with reliable many-body nuclear wave functions. 
Fortunately, in the case of the scalar
coupling, which is viewed as the most important, the situation is
a bit simpler, as only the nuclear form
factor is needed.
\item Knowledge of the WIMP density in our vicinity and its velocity
distribution. Since the essential input here comes from the rotation
curves, dark matter candidates other than the LSP are
also characterized by similar parameters.
\end{enumerate}

In the past various velocity distributions have been considered for
the dark matter gas in our galaxy. The most popular one is the
isothermal Maxwell-Boltzmann (M-B) velocity distribution with
$\langle\upsilon ^2\rangle=3v^2_d\simeq 3\upsilon_0^2/2$, where $v^2_d
=\langle v^2_x \rangle=\langle v^2_y \rangle=\langle v^2_z \rangle$
and $\upsilon_0$ is the velocity of the sun around the galaxy,
i.e. $\upsilon_0\simeq 220$ km/s.  Extensions of the M-B distribution
have also been considered, in particular those that are axially
symmetric with enhanced dispersion in the galactocentric direction
\cite{Druk,Verg00,evans00,fornen}. 
The need for such a velocity anisotropy is numerically well founded
\cite{HANSEN05}.  It has also been shown that the velocity anisotropy
is, in principle, measurable in a direction sensitive experiment
\cite{Host}, and it has possibly been observed to be non-zero in
clusters of galaxies~\cite{rocco}.

Non-isothermal models have also been considered. Among these one
should mention dark matter orbiting the Sun \cite{Krauss}, or dark
matter which is part of the Sagittarius tidal stream \cite{GREEN02}.
The velocity distribution has also been obtained in ``adiabatic''
models employing the Eddington method
\cite{EDDIN,UK01,VEROW06,Belli:2002yt}. In such an approach, given the
density of matter, one can obtain a distribution that depends both on
the velocity and the gravitational potential. Evaluating this
distribution in a given point in space, e.g. in our vicinity, yields
the velocity distribution at that point in a self-consistent manner.
Unfortunately this approach is applicable only if the density of
matter is spherically symmetric and the distribution depends only on
energy.  Also variants of the M-B distribution resulting from a
coupling of dark matter to dark energy \cite{TETRVER06} have been
considered.

In the present work we will consider a Tsallis type velocity
distribution \cite{TSALLIS88}. This has been found to be a good
description of the velocity distributions in numerical simulations
with realistic dark matter density and anisotropy profiles
\cite{HANSEN06}. We compare the direct dark matter detection rates
obtained in this way with the results for an axially symmetric
Maxwell-Boltzmann velocity distribution.

\section{A realistic velocity distribution}

Let us first introduce and discuss all the details of the velocity
distribution function (VDF) which we will use.

Very often the VDF of the dark matter particles is approximated by a
Maxwell-Boltzmann (M-B) shape (a Gaussian).  
There are many good reasons for doing this, in particular
the fact that many steps can be made analytically since the M-B
is easy to integrate. It is, however, well known that the M-B is
only an approximation, which is accurate only for an isothermal
sphere. For more general density profiles the shape of the VDF has a
different form. For example, when considering a power-law in density
over many orders of magnitude one finds for isotropic structures that
the VDF has the Tsallis shape~\cite{HANSEN04}. The Tsallis VDF depends
on a entropic index $q$ in such a way that the normal M-B is
the limiting case for $q\rightarrow 1$~\cite{TSALLIS88}. The shape of
the VDF for more realistic cosmological dark matter structures has
still not been calculated analytically, partly because the
distribution function cannot be assumed to depend only on
energy. First steps in the direction of actually deriving the VDF from
a generalized Eddington method have been taken~\cite{anevans}.

It has recently been identified that the {\em shape} of the VDF
actually is different for the galactic radial and tangential
directions \cite{HANSEN06}. This means that when the particle velocity
is decomposed into the radial and the tangential component with
respect to the galactic center, the corresponding distributions
are different. Specifically, it appears that the tangential VDF is always well fit
with the Tsallis shape using an entropic index of $q=5/3$, at least for
velocities smaller than 1.6 times the velocity dispersion.  Thus the
tangential VDF is universal at any radius and the only free
parameter is the tangential velocity dispersion.

The radial VDF, however, differs strongly as a function of radius. It
appears that the form of the radial VDF is fairly similar to what
comes out of the Eddington method, when using a density profile of the
NFW \cite{NFW} or Sersic shape \cite{grah02,ssole07}. 
In the inner part of the dark
matter halo the shape of the radial VDF is fairly simple, and again
the shape is reasonably well fit by the Tsallis shape with entropic
index fairly close to unity. A clear advantage of approximating the real
VDFs by distributions of the Tsallis shape is, as in the Maxwell-Boltzmann
case, that certain integrals can be done analytically.

We therefore consider a distribution function in the radial direction
which is given by: \beq\label{eq:fr}
f_r(\upsilon,\sigma_r,q)=N(q,\sigma_r)\left(1-\frac{(1-q)
\upsilon^2}{(3-q) \sigma_r^2}\right)^{\frac{q}{1-q}},
~~\upsilon^2=\upsilon_r^2.  \eeq where $\sigma_r$ is the radial
velocity dispersion and $N(q,\sigma_r)$ is a normalization factor. We
note that
$$\left(1-\frac{(1-q) \upsilon^2}{(3-q)
\sigma_r^2}\right)^{\frac{q}{1-q}}\rightarrow \exp\left(-\frac{v^2}{2
\sigma_r^2}\right)\quad \mbox{ as } q \rightarrow 1.$$ For $q>1$ this
function is defined on the whole axis. For $q<1$ the domain is
limited, since one must demand that the function is non-negative. Thus
for $q=3/4$ the function becomes
\beq\label{eq:fr1}
f_r(\upsilon,\sigma_r)=\frac{35}{96\sigma_r}\left
(1-\frac{\upsilon^2}{9 \sigma_r^2}\right )^3,\qquad -3 \sigma_r\leq
\upsilon\leq 3 \sigma_r \, .  \eeq We emphasize that this distribution
thus automatically imposes an upper bound on the acceptable
velocities. In the case of the M-B distribution this is imposed by
hand, $\upsilon_{\text{esc}}=2.84 \upsilon_0$, where $\upsilon_0$ is
the solar rotational velocity.
     \begin{figure}[bp]
 \begin{center} 
 \rotatebox{90}{\hspace{1.0cm} {$f_r(\upsilon)\sqrt{2 \sigma} \longrightarrow$}}
    \put(30,165){1-dim M-B ($\beta=0$)$\uparrow$ }
 \put(60,225){$\downarrow$ 1-dim M-B ($\beta=0.3$)}
 \includegraphics[scale=1.3]{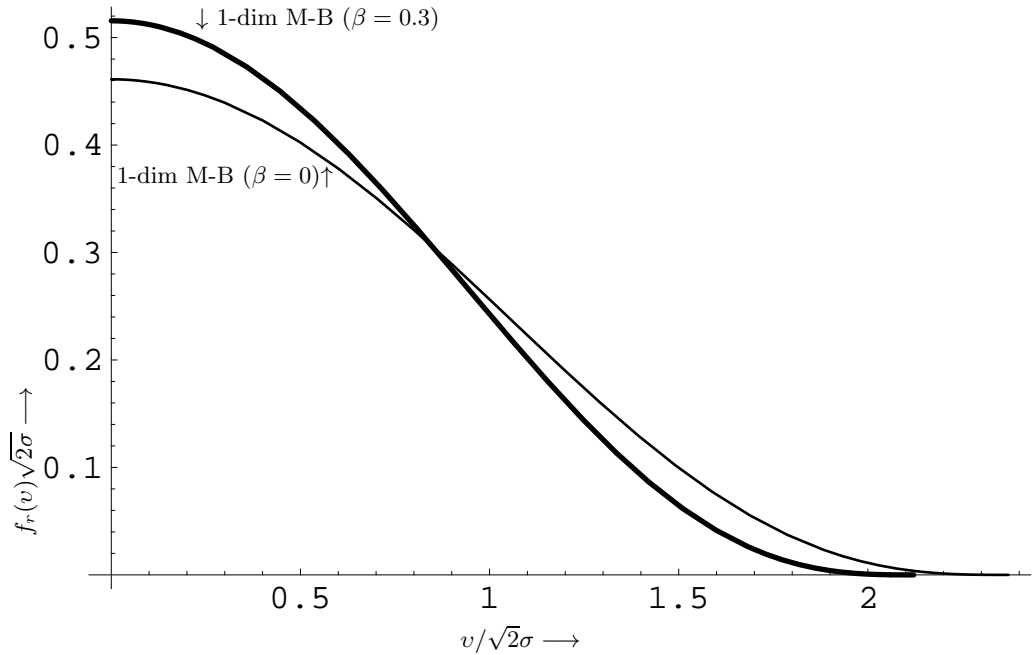}
 \hspace{-0.0cm} {$\upsilon/\sqrt{2} \sigma \longrightarrow$}
 \caption{The normalized velocity distribution in the radial direction $f_r(\upsilon)$ $(\upsilon=\upsilon_r)$
for $\beta=0$ as a function of $\upsilon/\sigma \sqrt{2}$
for $q=3/4$ (thin line) and the one dimensional M-B distribution (thick line). For $\beta=0$, $\sigma=\sigma_r$. 
Since this one dimensional distribution  is symmetric, only the sector $\upsilon\geq 0$ is shown.
}
 \label{fig:fr}
 \end{center}
  \end{figure} 

The corresponding Maxwell-Boltzmann distribution, i.e.~the limit of
\eqref{eq:fr} as $q\rightarrow1$, is given by:
\begin{equation}
f_{\text{MB}}(\upsilon,\sigma_r)=\frac{1}{\sqrt{2 \pi } \sigma_r}
\exp\left(-\frac{\upsilon^2}{2 \sigma_r^2} \right) \, . \end{equation}

Both the Tsallis shape with $q=3/4$ and the M-B shope (corresponding to $q=1$) 
are shown in Fig.~1.

The tangential velocity distribution, defined and normalized in two
dimensions, is characterized by $q>1$ and takes the form:
\begin{equation}
f_t(\upsilon,\sigma_t)=\frac{1}{2\pi\sigma_t^2(2-q)}\left (1-\frac{q-1}{2(q-2)}\left(\frac{\upsilon}{\sigma_t}\right)^2 \right)^{\frac{q}{1-q}},~~ \upsilon^2=\upsilon_t^2.
\end{equation}
Introducing the asymmetry parameter $\beta$ defined by 
\begin{equation}
\beta=1-\frac{\sigma^2 _t}{\sigma^2_r},
\end{equation}
and using the relation for the total velocity dispersion
\beq
3 \sigma^2=\sigma_r^2+2 \sigma_t^2
\eeq
we can express this distribution in terms of $\beta$ and $\sigma$. Thus we obtain:
\begin{equation}
f_t(\upsilon,\beta,\sigma)=\frac{1}{2\pi\sigma^2(2-q)}\frac{1-\frac{2}{3}\beta}{1-\beta}\left (1-\frac{(q-1)}{2(q-2)}
\frac{1-\frac{2}{3}\beta}{1-\beta}\left(\frac{\upsilon}{\sigma}\right)^2 \right)^{\frac{q}{1-q}},\qquad1<q< 2.
\label{Eq:ft}
\end{equation}

\begin{figure}[tbp]
 \begin{center}
  \rotatebox{90}{\hspace{1.0cm} {$2 \pi \upsilon f_t(\upsilon)\sqrt{2 \sigma} \longrightarrow$}}
  \put(60,230){$ \downarrow $2-dim Tsallis (q=5/3,$\beta=0$)}
    \put(55,200){$ \leftarrow   $2-dim Tsallis (q=5/3,$\beta=0.3$)}
      \put(100,170){$ \downarrow $2-dim M-B ($\beta=0$)}
    \put(190,80){$ \leftarrow   $2-dim M-B ($\beta=0.3$)}
 \includegraphics[scale=1.3]{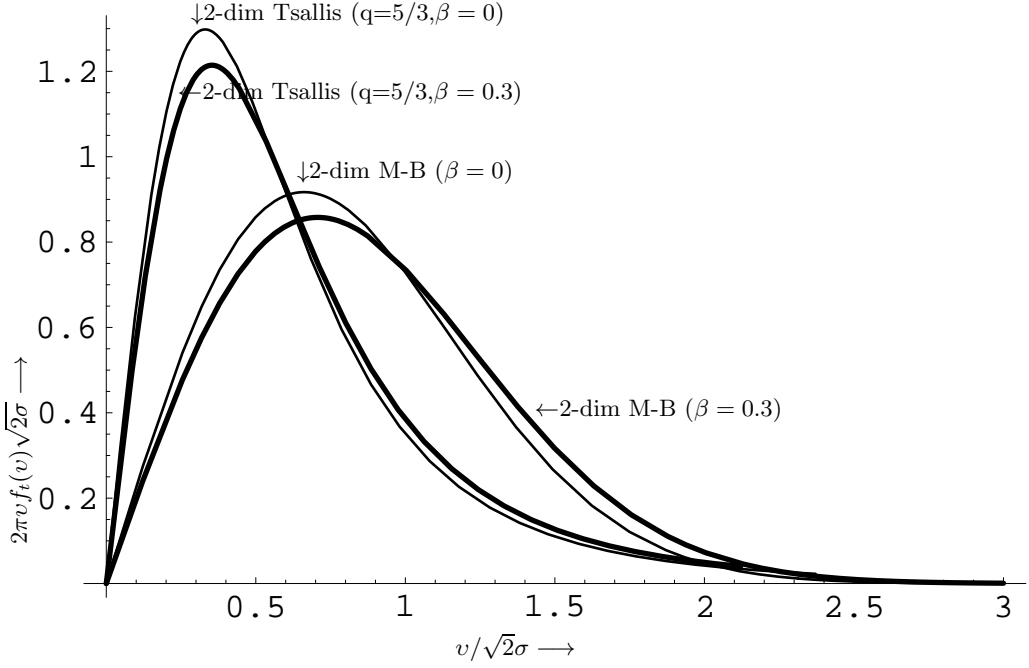}
  \hspace{-0.0cm} {$\upsilon/\sqrt{2} \sigma \longrightarrow$}
 \caption{The 2-dimensional Tsallis and M-B normalized tangential
 velocity distributions $2 \pi \upsilon f_t(\upsilon)$ are shown as
 functions of $\upsilon/\sigma \sqrt{2}$ ($\upsilon=\upsilon_t$), for
 values of the asymmetry parameter $\beta=0$ (thick line) and
 $\beta=0.3$ (fine line). The Tsallis curves with $q=5/3$ are the ones
 reaching the highest maximum value at lower velocity.}
 \label{fig:vftboth}
 \end{center}
  \end{figure}

The corresponding two-dimensional M-B distribution with asymmetry $\beta$ is given by:
\begin{equation}
f_{\text{MB}}(\upsilon,\beta,\sigma)= \frac{1-\frac{2}{3}\beta}{2 \pi  (1-\beta ) \sigma^2}
\exp\left(-\frac{(1-\frac{2}{3}\beta)\upsilon ^2}{2 (1-\beta ) \sigma ^2}\right).   
\end{equation}
This can be obtained from eq.~(\ref{Eq:ft}) in the limit $q\rightarrow
1$.  As explained earlier, we assumed $q=3/4$ for $f_r$ and $q=5/3$
for $f_t$ for the more realistic VDF, which implies that the range of allowed velocities is
$\upsilon<3 \sigma/\sqrt{1-(2/3)\beta}$, which is slightly different
from the relation $\upsilon<2.84 \upsilon_0 \approx 4 \sigma$
normally imposed on the M-B distribution. The shapes of the two
2-dimensional tangential velocity distributions are shown together in
Fig.~\ref{fig:vftboth} and exhibit substantial differences in the
width and the attained maximum, as well as at high velocities. Note that the
factor $2\pi v$ is just from the 2-dimensional phase-space volume.

In terms of $\beta$ and $\sigma$ the radial velocity distributions are
\beq
f_r(\upsilon,\beta,\sigma)= \frac{35 \sqrt{1-\frac{2}{3} \beta }
\left(1-\frac{\upsilon^2 \left(1-\frac{2 \beta }{3}\right)}{9 \sigma
^2}\right)^3}{96 \sigma } \mbox{ (Tsallis function, $q=3/4$) }, \eeq
and
\begin{equation}
f_r(\upsilon,\beta,\sigma)=
\frac{e^{-\frac{\upsilon^2 \left(1-\frac{2 \beta }{3}\right)}{2 \sigma ^2}}
   \sqrt{1-\frac{2 \beta }{3}}}{ \sqrt{2 \pi} \sigma }
 \mbox{  (MB) }.
\end{equation}

We note that in the case of the M-B distribution
$\sigma=\upsilon_0/\sqrt{2}$, where $\upsilon_0$ is the sun's
rotational velocity. In the case of the Tsallis functions considered
here, one has $\langle\upsilon^2\rangle=3 \sigma^2$ for all q. We impose the
condition $\langle\upsilon^2\rangle=(3/2)\upsilon_0^2$ (as for the M-B
distribution) even for $q\ne1$ which implies $\sigma=
\upsilon_0/\sqrt{2}$.

\section{The direct detection event rate}

We will now calculate both the total detection rate and the annual 
modulation rate. However,  
before computing the detection event rate it is necessary to transform
the distributions from the galactic to the local coordinate
system.

From the kinematics of the WIMP-nucleus collision one has that the
momentum transfer to the nucleus is given by \beq q=2 \mu_r \upsilon
\xi, \eeq where $\xi$ is the cosine of the angle between the WIMP
velocity and the momentum of the outgoing nucleus, and $\mu_r$ is the
reduced mass of the system. Instead of the variable $\xi$ one can
introduce the energy $Q$ transferred to the nucleus, $Q={q^2}/({2 A
m_p})$, where $A m_p$ is the nuclear mass. Thus
$$2 \xi d\xi=\frac{A m_p}{2 (\mu_r \upsilon)^2} dQ.$$
Furthermore, for a given energy transfer the velocity $\upsilon$ is constrained to be
\beq
\upsilon\geq \upsilon_{min}~,~\upsilon_{min}= \sqrt{\frac{ Q A m_p}{2}}\frac{1}{\mu_r}.
\eeq
We will find it convenient to introduce, instead of the energy transfer, the dimensionless quantity $u$
\beq
u=\frac{1}{2}(qb)^2\equiv\frac{Q}{Q_0}~~,~~Q_{0}=\frac{1}{Am_p b^2}=4.1\times 10^{4}~A^{-4/3}~{\rm keV},
\label{u.1}
\eeq  
where $b$ is the nuclear (harmonic oscillator) size parameter. 

It is therefore clear that for a given energy transfer the velocity is
restricted from below, and we have already mentioned that the velocity
is bounded from above by the escape velocity.  We introduce a
normalized velocity by the dimensionless variable
\begin{equation}
y=\upsilon/(\sigma \sqrt{2}),\qquad a \sqrt{u}\leq y\leq y_{\text{esc}},
\end{equation}
with $a= \left(2\mu_r b \sigma \right)^{-1}$.
Thus,
\beq
\xi d\xi=\frac{a^2}{y^2} dy.
\label{utrans}
\eeq

The event rate for the coherent WIMP-nucleus elastic scattering is given by 
\cite{Verg01}:
\beq
R= \frac{\rho (0)}{m_{\chi^0}} \frac{m}{m_p}~
              \sqrt{\langle v^2 \rangle } f_{\text{coh}}(A,\mu_r(A)) \sigma_{p,\chi^0}^{S},
\label{fullrate1}
\eeq
or, using typical numerical values,
\beq
 R\simeq 1.60~10^{-3}
\frac{t}{1 \mbox{y}} \frac{\rho(0)}{0.3\mbox{GeVcm}^{-3}}
\frac{m}{\mbox{1kg}}\frac{ \sqrt{\langle
v^2 \rangle }}{280 \mbox{kms}^{-1}}\frac{\sigma_{p,\chi^0}^{S}}{10^{-6} \mbox{ pb}} f_{\text{coh}}(A, \mu_r(A)).
\label{eventrate}
\eeq
Here, $\sigma_{p,\chi^0}^{S}$ is the WIMP-nucleon scalar cross section, $\rho(0)$ is the WIMP density in our vicinity, $m_{\chi^0}$ is the WIMP mass, $m$ is the target mass, $A$ is the number of nucleons
in the nucleus, $\langle v^2 \rangle=3v_0^2/2$ is the root-mean-square WIMP velocity. The dimensionless quantity 
$f_{\text{coh}}(A, \mu_r(A))$ is given by
\beq
f_{\text{coh}}(A, \mu_r(A))=\frac{100\mbox{GeV}}{m_{\chi^0}}\left[ \frac{\mu_r(A)}{\mu_r(p)} \right]^2 A~t_{\text{coh}}\left(1+h_{\text{coh}}\cos\alpha \right).
\label{fullrate2}
\eeq
The quantity of interest to us is $r=
t_{\text{coh}}\left(1+h_{\text{coh}}\cos\alpha \right)$, which is also
dimensionless. 
The angle $\alpha$ describes the position of the Earth in its solar orbit.
This can be viewed as a "relative" rate and contains
all the information regarding the WIMP velocity distribution and the
structure of the nucleus. It also depends on the reduced mass of the
system.

The event rate is proportional to the WIMP flux, i.e.~proportional to
the WIMP velocity.  
It is not difficult to show
\cite{Verg01} that \beq \frac{\mathrm{d}r}{\mathrm{d}u}= F^2(u)
\int_{a \sqrt{u}}^{y_{\text{esc}}} \sqrt{\frac{2}{3}} y
\frac{a^2}{y^2} y^2 dy \int_{0}^{\pi} d \theta \int_{0}^{2 \pi} d \phi
f_r(y,\theta)f_t(y,\theta,\phi, \alpha) \Theta(y_{\text{esc}}^2-f_b),
\label{Eq:drdu}
\eeq with $F(u)$ the nuclear form factor. 
In the integrand we
have explicitly displayed all the factors of $y$ in order to keep
track of their origin. The first one comes from the WIMP flux, the second
from the transformation (\ref{utrans}) and the last is the usual
phase-space factor. $\Theta(x)$ is the Heavyside function introduced to
guarantee that the WIMP velocity is bounded by the escape velocity
Thus,
$$
f_b= y^2 + y_0^2 + \delta_0^2 + y_0 \delta_0 \cos{\alpha} + 
    2 y \delta_0  \cos{\theta} \sin{\alpha}$$
 Even though the values of $y_{esc}$ are different, the effect of this
 constraint is small, with the possible exception of the event rates
 for very high threshold energy. In that special case the difference
 in the upper cutoff, $y_{\text{esc}}$, between the two models may
 yield significant differences.

 This way the differential event rate in eq.~\eqref{Eq:drdu} can be cast in the form:
\begin{equation}
\frac{dr}{du}=\sqrt{\frac{2}{3}}a^2 F^2(u)  \Psi(a \sqrt{u},\alpha ). 
\label{eq:rrateall}
\end{equation}
We have seen that the parameter $a$ depends on the nucleus, the
velocity distribution ($\upsilon_0$ or $\sigma$) and the WIMP mass.

By performing a Fourier analysis of the function $\Psi(x,\alpha)$
which is a periodic function of $\alpha$ and keeping the dominant
terms we find:
\begin{equation}
\frac{dr}{du}=\sqrt{\frac{2}{3}} a^2 F^2(u) \Psi_0(a \sqrt{u})\left [1+H(a \sqrt{u}) \cos{\alpha} \right ].
\label{eq:rrate}
\end{equation}
Sometimes we will consider each term separately in the above expression by writing:
\beq
\frac{dr}{du}= \frac{dt}{du}+\frac{d \tilde{h}}{du}\cos{\alpha}~,~\frac{dt}{du}=\sqrt{\frac{2}{3}} a^2 F^2(u) \Psi_0(a \sqrt{u})~,~\frac{d \tilde{h}}{du}=\sqrt{\frac{2}{3}} a^2 F^2(u) \Psi_0(a \sqrt{u})H(a \sqrt{u})
\eeq

It is thus clear, that $\Psi_0$ and $t$ are related to the total
rates, whereas $H$ and $h$ are related to the annual modulations.

\begin{figure}[t]
\begin{center}
\subfloat[]
 {
  \rotatebox{90}{\hspace{-0.0cm} {$\Psi_0(a\sqrt{u}) \longrightarrow$}}
\includegraphics[clip,width=0.4\linewidth]{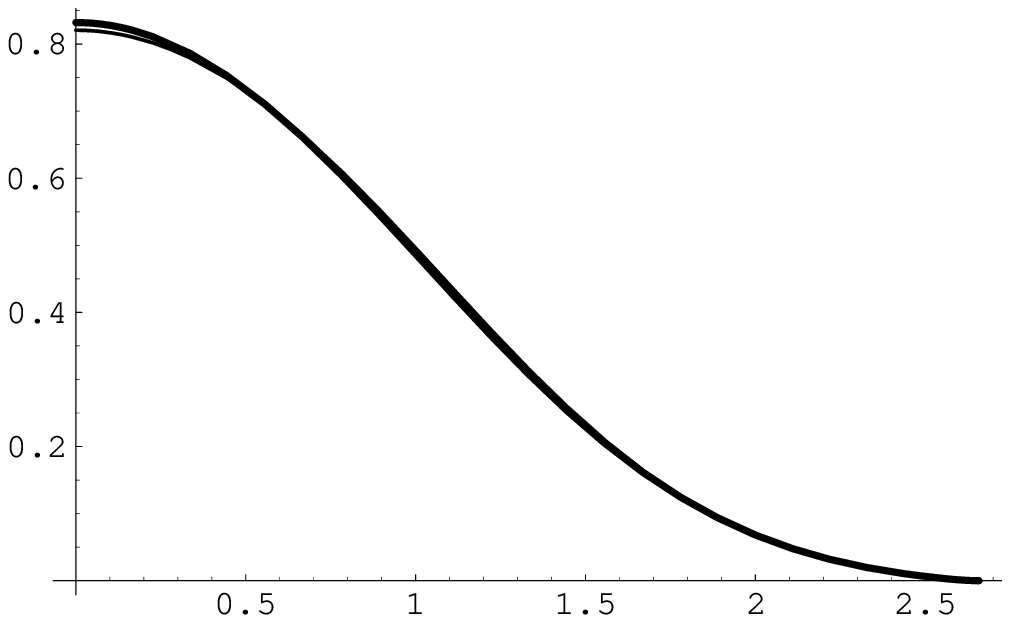}
}
\subfloat[]
 {
 \rotatebox{90}{\hspace{-0.0cm} {$\Psi_0(a\sqrt{u}) \longrightarrow$}}
\includegraphics[clip,width=0.4\linewidth]{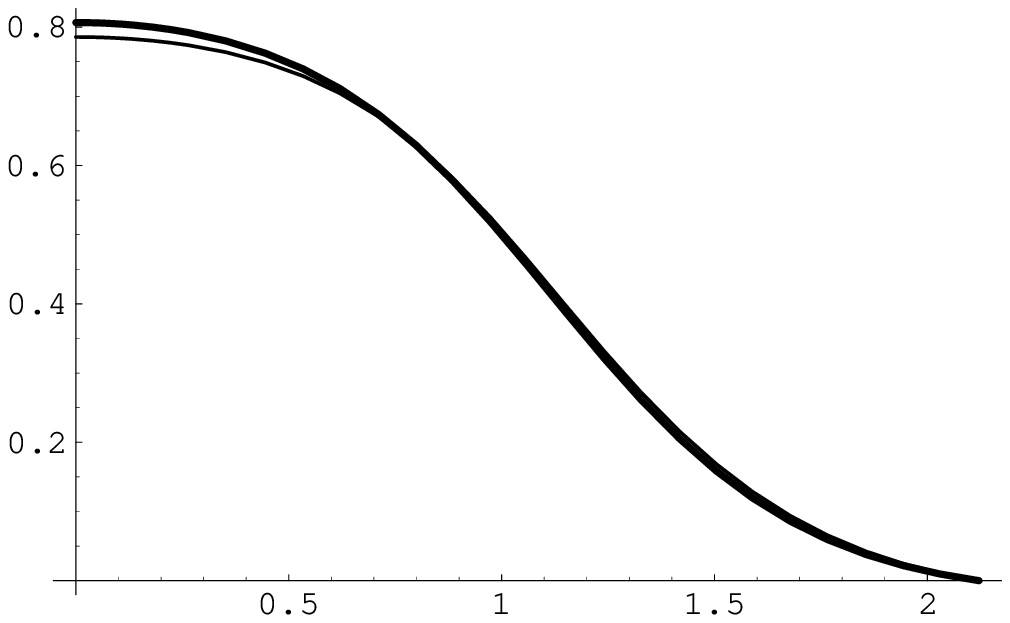}
}\\
\hspace{-0.0cm} {$a \sqrt{u} \longrightarrow$}
\caption{The time independent part of the differential rate, $\Psi_0(
a \sqrt{u})$, as a function of the energy transfer $a\sqrt{u}$.  The normally
used M-B distribution is on the left and the present realistic velocity
distribution is on the right.
There is essentially no  dependence on the velocity asymmetry parameter.  }
 \label{fig:nomod}
 \end{center} 
  \end{figure}

\begin{figure}[tbp]
 \begin{center}
 \subfloat[]
 {
  \rotatebox{90}{\hspace{-0.0cm} {$H( a \sqrt{u}) \longrightarrow$}}
 \includegraphics[clip,width=0.4\linewidth]{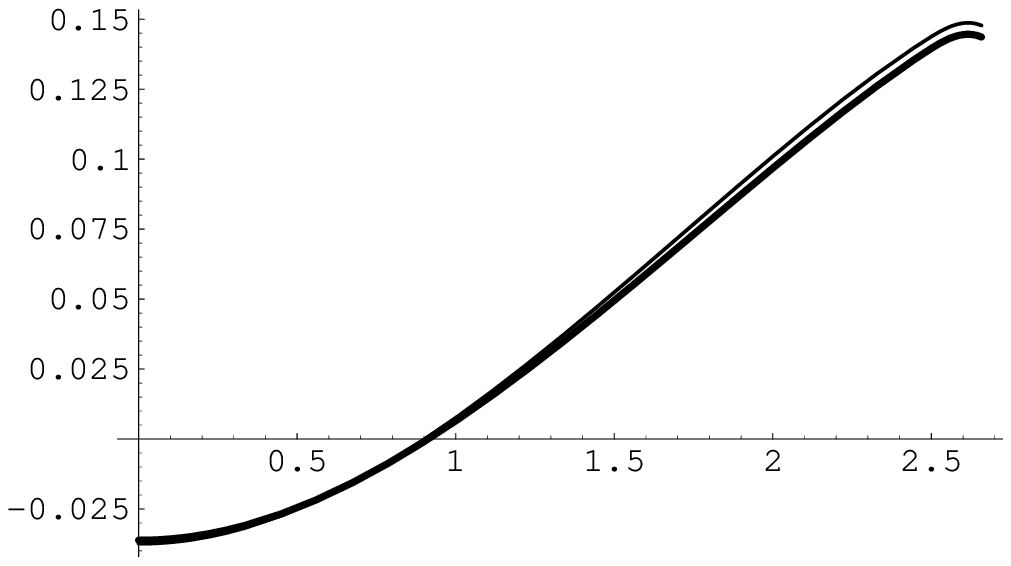}
 }
\subfloat[]
 {
  \rotatebox{90}{\hspace{-0.0cm} {$H(x) \longrightarrow$}}
  \includegraphics[clip,width=0.4\linewidth]{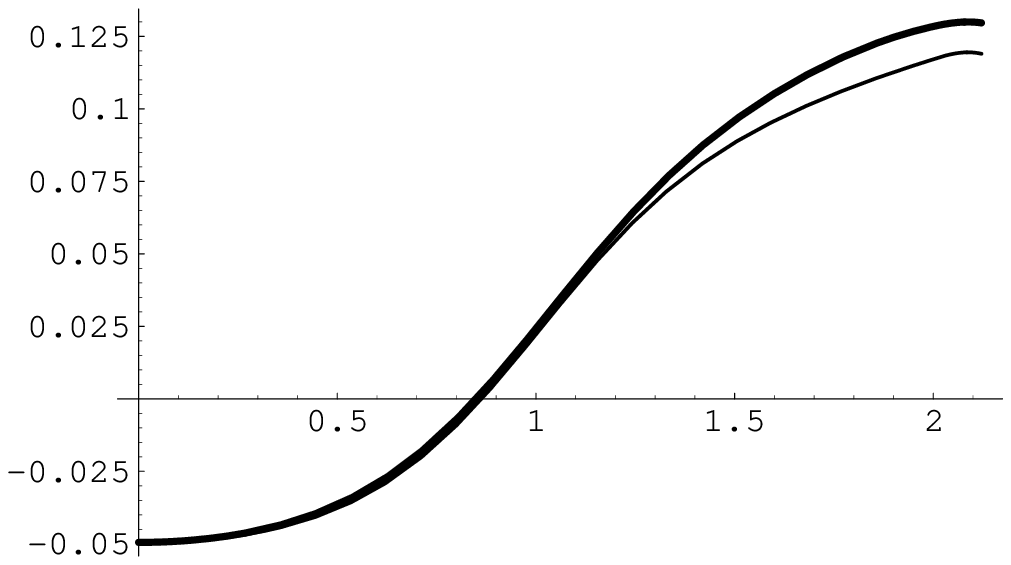}
  }\\
  \hspace{-0.0cm} {$ a \sqrt{u} \longrightarrow$}
 \caption{The same as in Fig.~\ref{fig:nomod} for the function $H( a
 \sqrt{u})$, which describes the coefficient of the time dependent
 part of the differential rate (differential modulated amplitude).
 Note the change in sign as $ a \sqrt{u}$, i.e.  the energy transfer,
 increases. This leads to cancellations in the total modulation
 amplitude. M-B is on the left, Tsallis is on the right.
In some cases it may even become negative (maximum event
 rate in December). The dependence on the asymmetry is small, visible
 only at high energy transfers.}
 \label{fig:mod} 
 \end{center}
  \end{figure}

Before proceeding further by considering a special target, it is
instructive to concentrate on $\Psi_0(x)$ and the modulation $H(x)$ as
functions of the asymmetry parameter $\beta$. For this purpose we show
the function $\Psi_0(x)$ in Fig.~\ref{fig:nomod} and the function
$H(x)$ in Fig.~\ref{fig:mod}.
In the case of $\Psi_0( a \sqrt{u})$, one clearly sees that the effect
of the asymmetry parameter $\beta$ is small for all energy
transfers. In the case of the M-B distribution one has that larger  $\beta$
imply smaller differential rate, while in the case of the more realistic 
distribution a larger asymmetry implies a larger differential
rate.
 In both cases the function $H( a \sqrt{u})$ is an increasing function
of $a \sqrt{u}$ accompanied by a change in sign.  Thus in obtaining
the modulation of the total amplitude the contribution of the low $Q$
section tends to cancel the high $Q$ part. Which part will dominate
depends on the reduced mass and the nuclear form factor. In the
case of the present realistic distribution, the velocity asymmetry
has a small effect on the modulation of the differential rate,
and this effect occurs at high energy transfers.

\section{Applications}

Let us now focus on the aspects affected by the WIMP velocity distribution.
The total (time averaged) rate is given by:
 \beq
 t_{\text{coh}}=\int_{u_{\text{min}}}^{u_{\text{max}}} \frac{dt_{\text{coh}}}{du} du,
 \label{tcoh}
 \eeq
 where 
$u_{\text{min}}$ is determined by the detector threshold and 
$u_{\text{max}}=(y_{\text{esc}})^2/a^2$ by the maximum WIMP velocity.
By including  both $\Psi_0(a \sqrt{u})$ and $H(a \sqrt{u})$ we can cast the rate in the form:
\begin{eqnarray}
r_{\text{coh}}&=&t_{\text{coh}} \left(1+h_{\text{coh}} \cos{\alpha} \right)
\nonumber \\
h_{\text{coh}}&=&\frac{1}{t_{\text{coh}}} \int_{u_{\text{min}}}^{u_{\text{max}}} \frac{d\tilde{h}_{\text{coh}}}{du} du \, ,
\label{Eq:h}
\end{eqnarray}
where the dimensionless quantity $r$  characterizes the total  rate, and $h$ is the annual modulation.

The direct WIMP detection rate depends on the nucleus via its form
factor and its mass. It also depends on the WIMP mass through the
reduced mass $\mu_r$, entering through the parameter $a$.  For our
numerical study we will 
focus 
on two very different targets to show general trends,
namely
a medium heavy target, $^{127}$I, and
and a light one, $^{19}F$, which are two of the most popular targets
employed.
  \subsection{The case of a light target}
\begin{figure}[tbp]
 \begin{center}
  \subfloat[]
 {
\rotatebox{90}{\hspace{-0.0cm} {$F^2(u) \longrightarrow$}}
\includegraphics[scale=0.4]{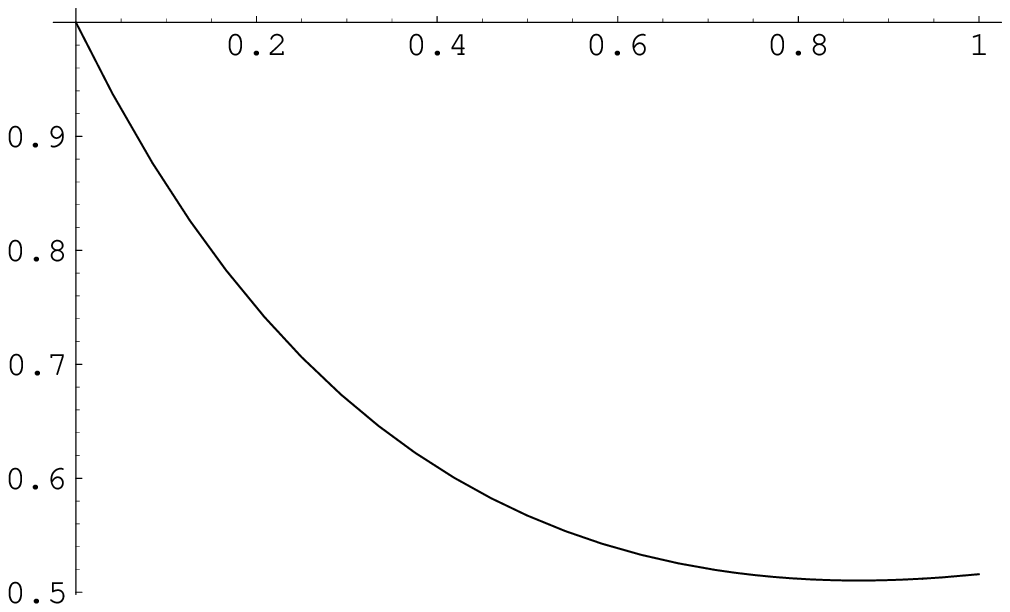}
{\hspace{0.0cm} $\longrightarrow  u$}
}
 \hspace{1.0cm}
 \subfloat[]
 {
\rotatebox{90}{\hspace{-0.0cm} {$F^2(Q) \longrightarrow$}}
\includegraphics[scale=0.4]{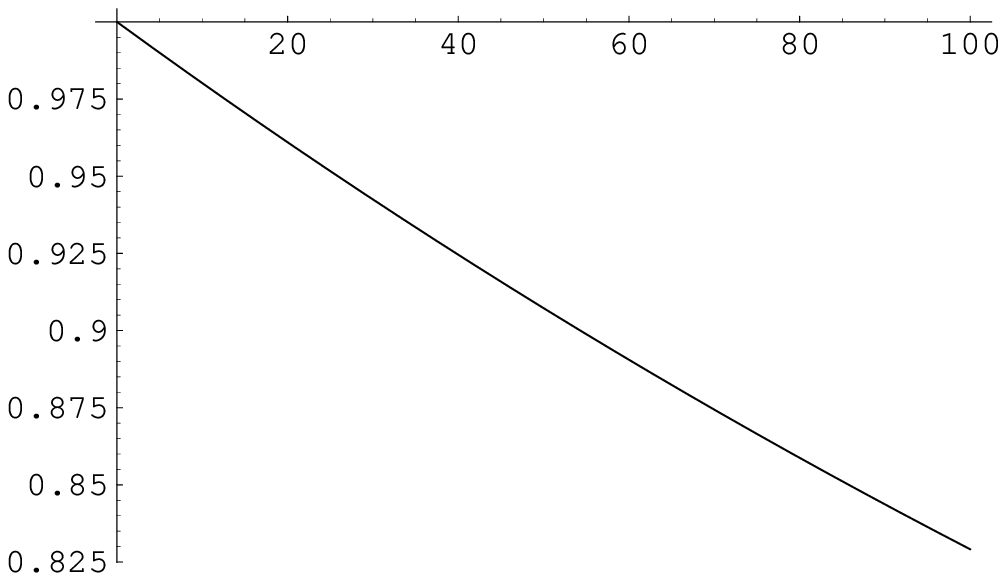}
{\hspace{0.0cm} $Q \longrightarrow  $ keV}
}
 \caption{(a) The form factor $F^2(u)$ for $^{19}$F employed in our
 calculation as a function of $u=Q/Q_0$, where Q is the energy transfer to the nucleus and
 $Q_0=809$ keV. (b) The same quantity as a function of the energy
 transfer $Q$.}
  \label{fig:sqformf19}
   \end{center}
  \end{figure}

\begin{figure}[tbp]
 \begin{center}
  \subfloat[]
 {
 \rotatebox{90}{\hspace{-0.0cm} {$t_{coh} \longrightarrow$}}
      \put(100,25){ $\downarrow \beta=0$}
          \put(100,2){ $\uparrow  \beta=0.3$}
 \includegraphics[clip,width=0.4\linewidth]{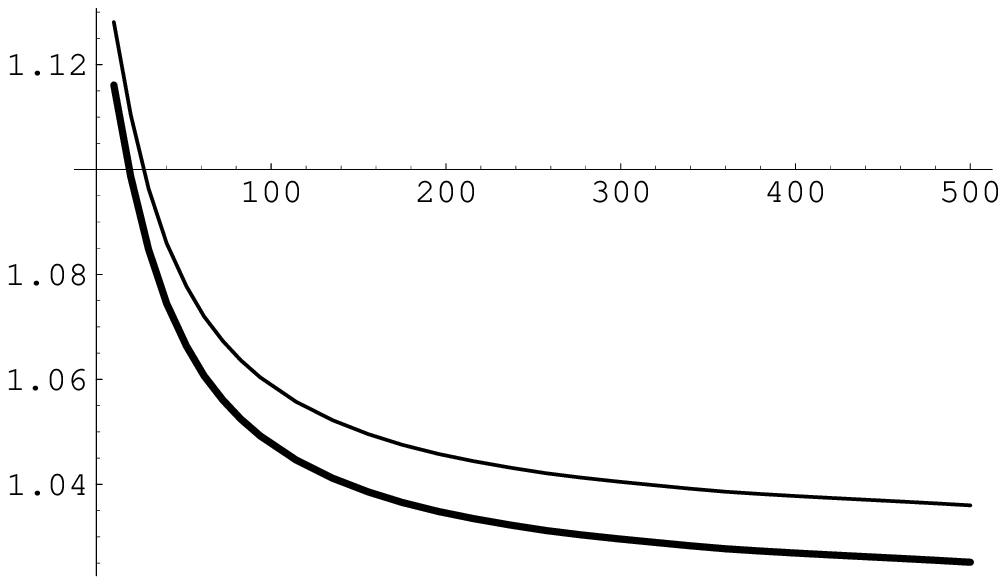}
 }
  \subfloat[]
 {
  \rotatebox{90}{\hspace{-0.0cm} {$t_{coh} \longrightarrow$}}
        \put(100,43){ $\downarrow \beta=0$}
          \put(100,17){ $\downarrow  \beta=0.3$}
  \includegraphics[clip,width=0.4\linewidth]{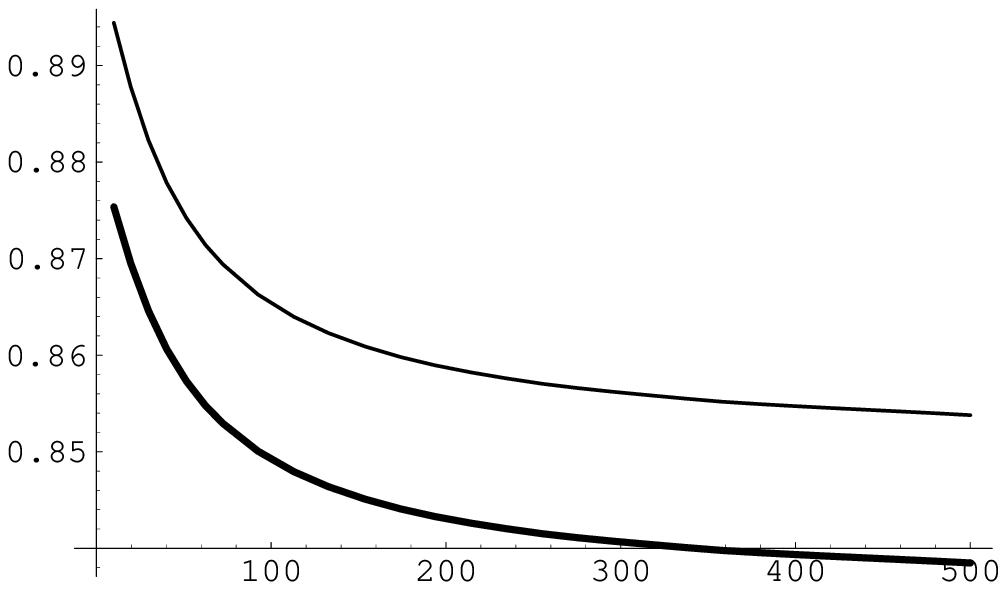}
}\\
    \hspace{-0.0cm} {$m_{\chi} \longrightarrow$ GeV}\\
     \subfloat[]
 {
  \rotatebox{90}{\hspace{-0.0cm} {$t_{coh} \longrightarrow$}}
          \put(100,93){ $\uparrow  \beta=0$}
          \put(100,70){ $\uparrow   \beta=0.3$}
   \includegraphics[clip,width=0.4\linewidth]{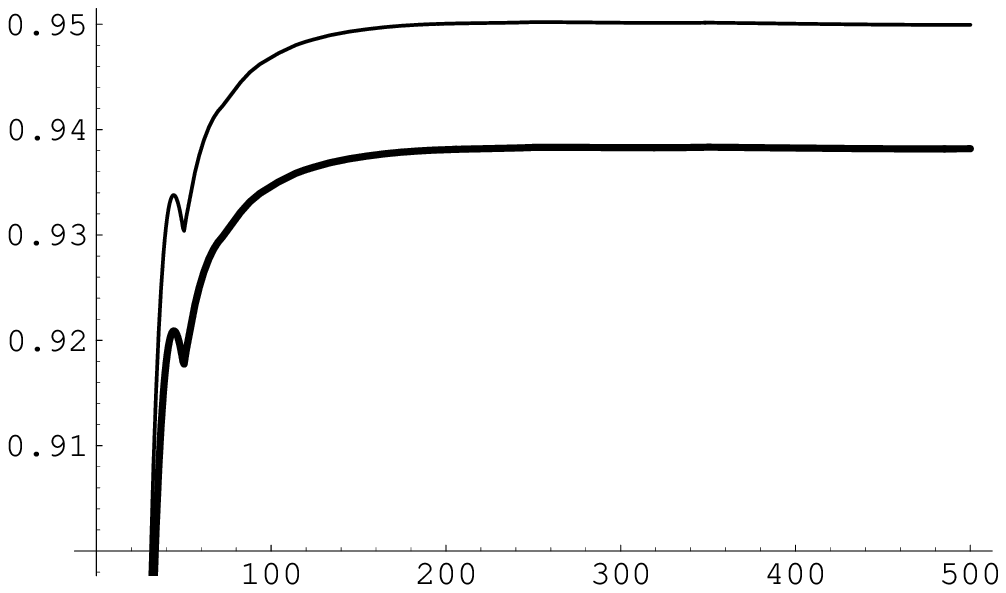}
   }
    \subfloat[]
 {
  \rotatebox{90}{\hspace{-0.0cm} {$t_{coh} \longrightarrow$}}
        \put(100,105){ $\downarrow   \beta=0$}
          \put(100,78){ $\uparrow   \beta=0.3$}
  \includegraphics[clip,width=0.4\linewidth]{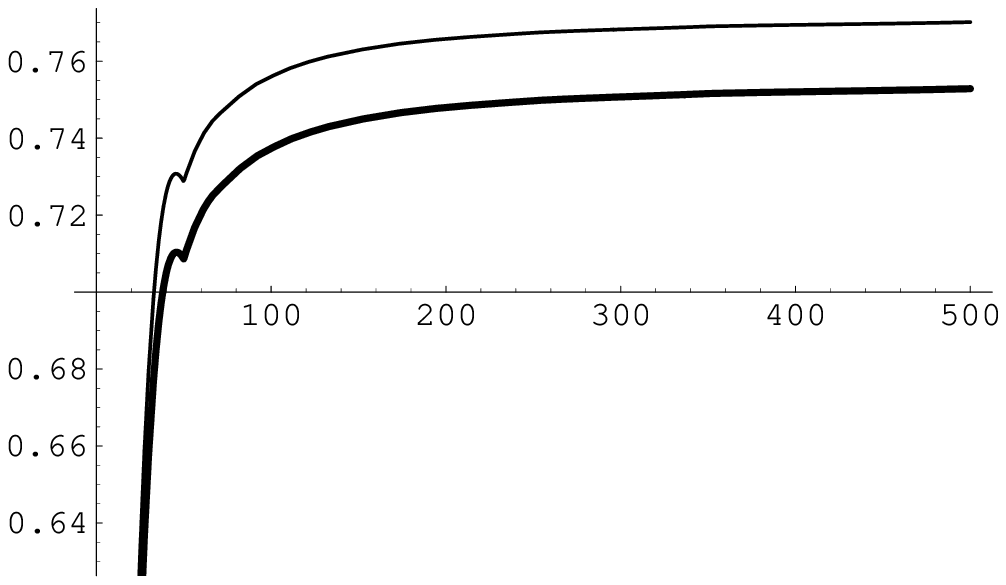}
 }\\
 \hspace{-0.0cm} {$m_{\chi} \longrightarrow$ GeV}
 \caption{The dimensionless quantity $t_{\text{coh}}$ in the case of the target
$^{19}$F for $Q_{\text{thr}}=0$ at the top and $Q_{\text{min}}=5$ keV
at the bottom. 
The abscissa is for different WIMP masses, $m_\chi$.
$t_{\text{coh}}$ represents the time-independent
part of the detection rate.
On the left the results obtained for a M-B
distribution, while on the right those for the more realistic 
distribution. The thick and thin curves correspond to $\beta=0$ and
$0.3$ respectively.
}
 \label{fig:Ftotalt}
 \end{center} 
  \end{figure}

\begin{figure}[tbp]
 \begin{center}
   \subfloat[]
 {
 \rotatebox{90}{\hspace{-0.0cm} {$h_{coh} \longrightarrow$}}
            \put(100,50){ $\downarrow   \beta=0$}
          \put(100,20){ $\downarrow    \beta=0.3$}
 \includegraphics[clip,width=0.4\linewidth]{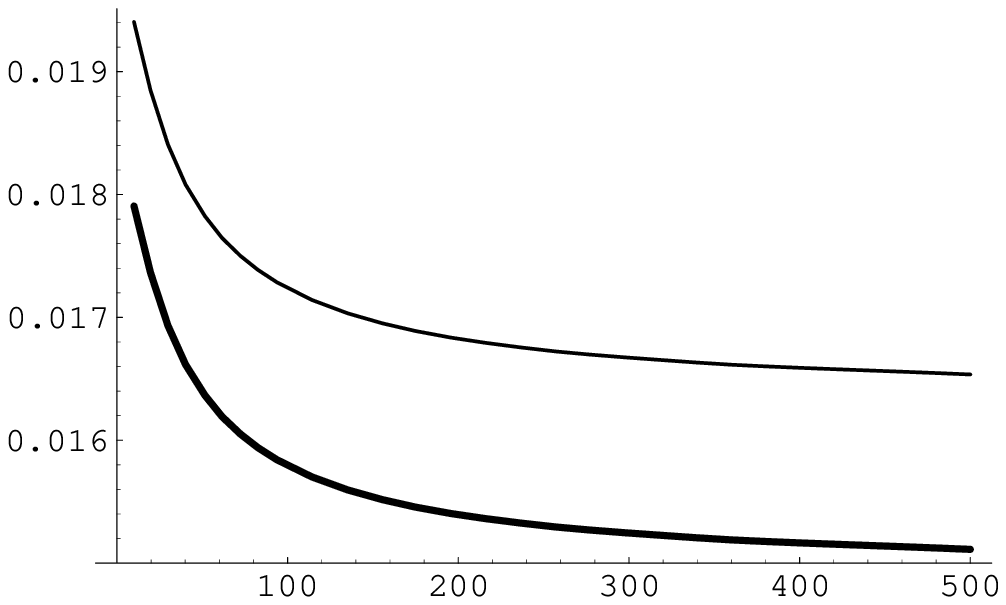}
 }
  \subfloat[]
 {
  \rotatebox{90}{\hspace{-0.0cm} {$h_{coh} \longrightarrow$}}
              \put(100,18){ $\downarrow   \beta=0$}
          \put(100,75){ $\downarrow    \beta=0.3$}
  \includegraphics[clip,width=0.4\linewidth]{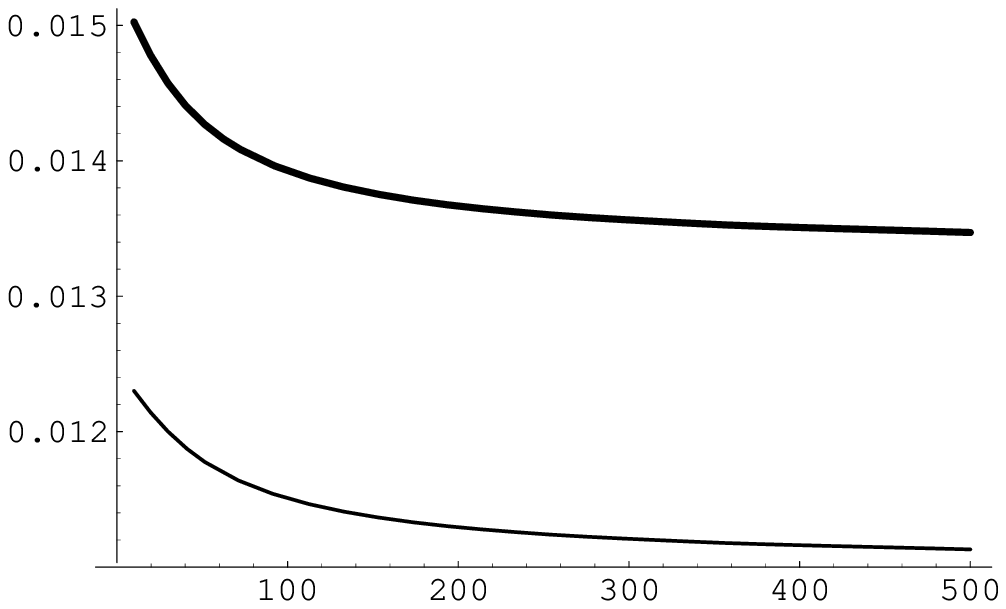}
}\\
    \hspace{-0.0cm} {$m_{\chi} \longrightarrow$ GeV}\\
     \subfloat[]
 {
  \rotatebox{90}{\hspace{-0.0cm} {$h_{coh} \longrightarrow$}}
              \put(100,15){ $\downarrow   \beta=0.3$}
          \put(100,25){ $\downarrow    \beta=0$}
   \includegraphics[clip,width=0.4\linewidth]{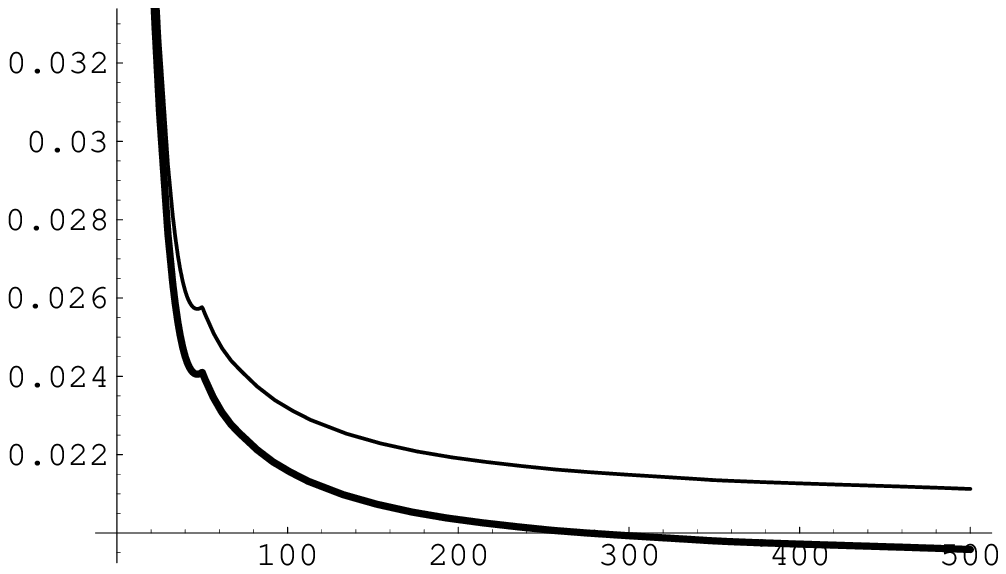}
   }
    \subfloat[]
 {
  \rotatebox{90}{\hspace{-0.0cm} {$h_{coh} \longrightarrow$}}
           \put(60,35){ $\downarrow   \beta=0.3$}
          \put(100,0){  $\uparrow   \beta=0$}
  \includegraphics[clip,width=0.4\linewidth]{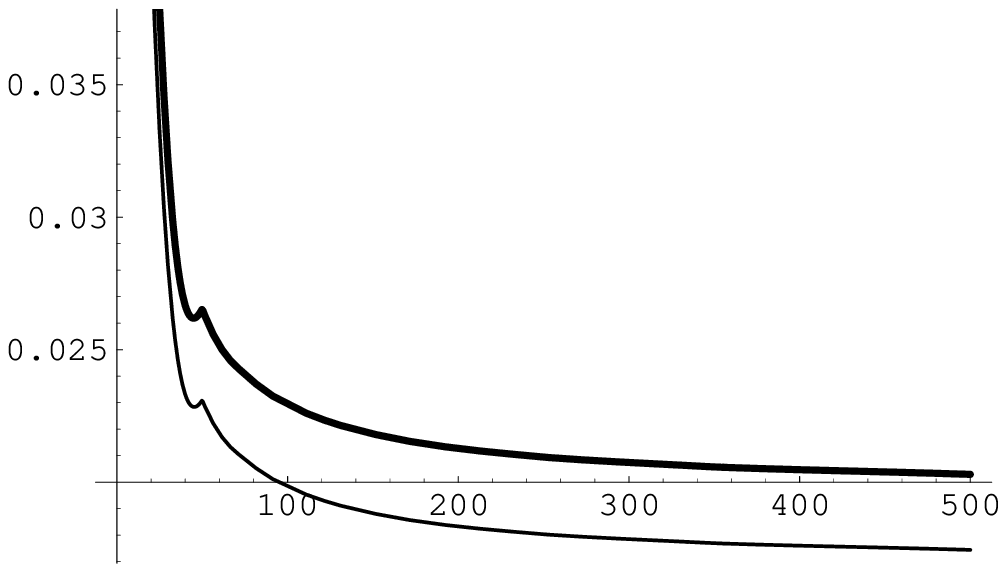}
 }\\
 \hspace{-0.0cm} {$m_{\chi} \longrightarrow$ GeV}
 \caption{The same as in Fig. \ref{fig:Ftotalt} for the quantity
$h_{coh}$, which represents the time-dependent part of the detection
rate (the annual modulation).
}
 \label{fig:Ftotalh}
 \end{center} 
  \end{figure}

The actual results have been obtained for  $^{19}$F, but those of other light targets are similar. The nuclear form factor we use was obtained in the shell model description of the target and is shown in fig.~\ref{fig:sqformf19}.
  Integrating over the energy transfer, assuming either no detector cut off ($u_{min}=0$) or a cut off of $Q_{\text{thr}}=5$ keV, we obtain the total rate in the case of the target $^{19}$F. The results are shown in Fig. \ref{fig:Ftotalt} 
for $t$ and in Fig. \ref{fig:Ftotalh} for $h$.
\subsection{The case of an intermediate mass target}
\begin{figure}[tbp]
 \begin{center}
  \subfloat[]
 {
\rotatebox{90}{\hspace{-0.0cm} {$F^2(u) \longrightarrow$}}
\includegraphics[scale=0.4]{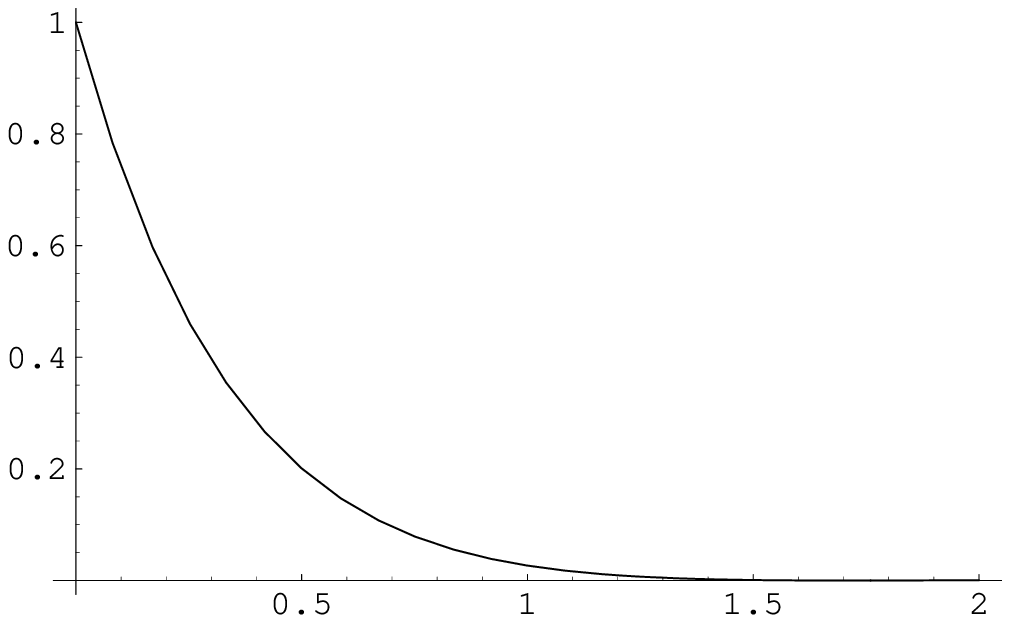}
{\hspace{0.0cm} $\longrightarrow  u$}
}
 \hspace{1.0cm}
 \subfloat[]
 {
\rotatebox{90}{\hspace{-0.0cm} {$F^2(Q) \longrightarrow$}}
\includegraphics[scale=0.4]{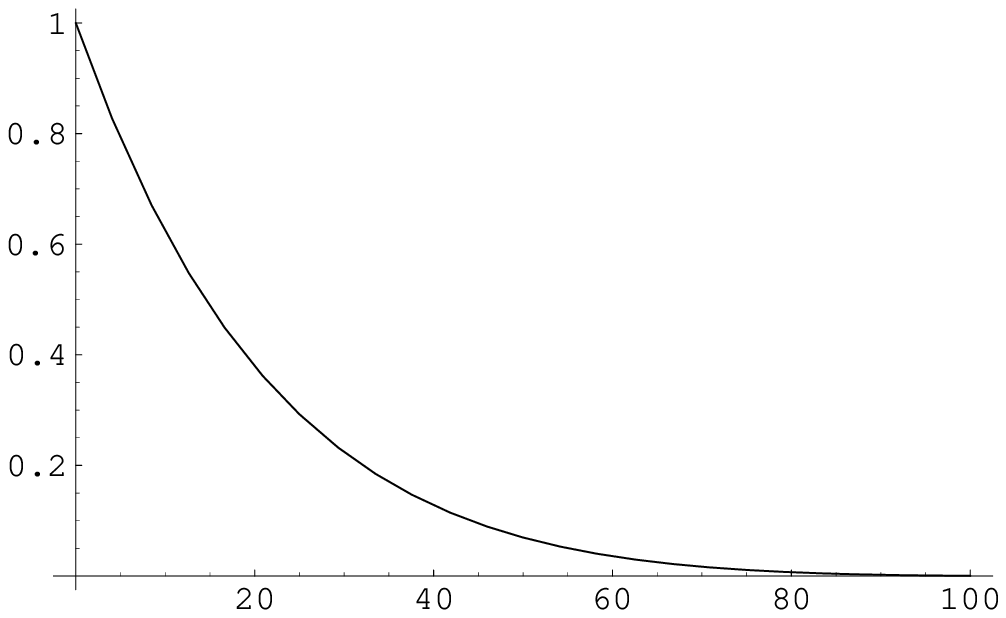}
{\hspace{0.0cm} $Q \longrightarrow  $ keV}
}
 \caption{(a) The form factor $F^2(u)$ for $^{127}$I employed in our
 calculation as a function of $u=Q/Q_0$, where Q is the energy transfer to the nucleus and
 $Q_0=64$ keV. (b) The same quantity as a function of the energy transfer
 $Q$.}
 \label{sqformf127}
   \end{center}
  \end{figure}

\begin{figure}[tbp]
 \begin{center}
   \subfloat[]
 {
 \rotatebox{90}{\hspace{-0.0cm} {$t_{coh} \longrightarrow$}}
 \includegraphics[clip,width=0.4\linewidth]{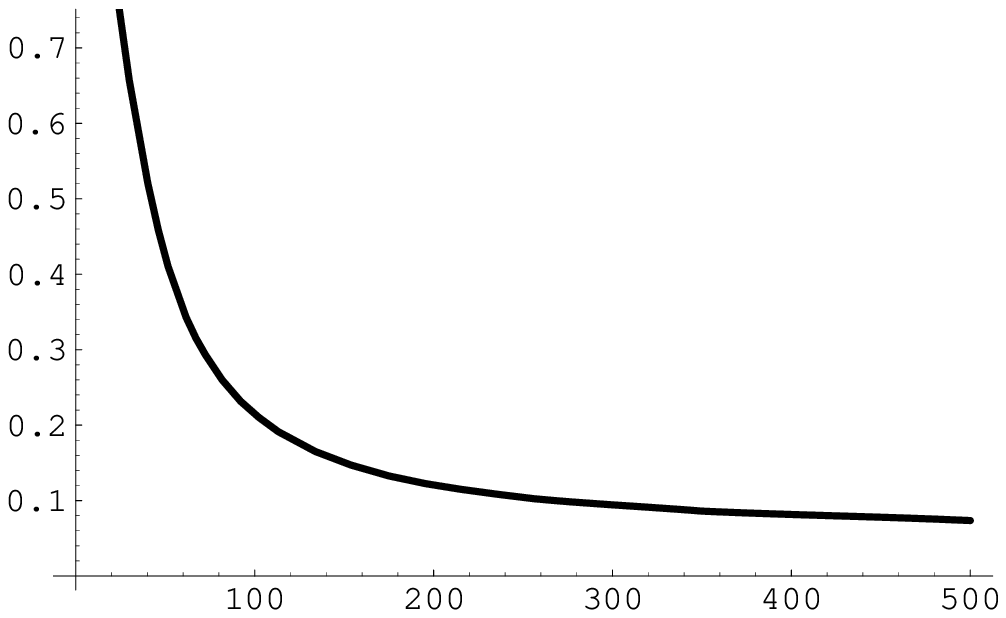}
 }
  \subfloat[]
 {
  \rotatebox{90}{\hspace{-0.0cm} {$t_{coh} \longrightarrow$}}
  \includegraphics[clip,width=0.4\linewidth]{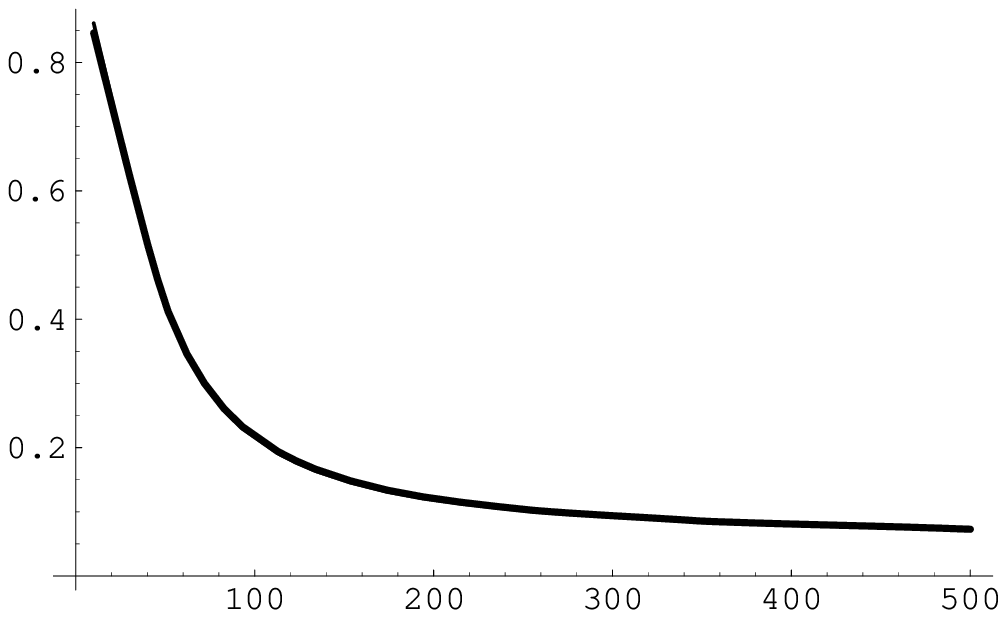}
}\\
    \hspace{-0.0cm} {$m_{\chi} \longrightarrow$ GeV}\\
     \subfloat[]
 {
  \rotatebox{90}{\hspace{-0.0cm} {$t_{coh} \longrightarrow$}}
   \includegraphics[clip,width=0.4\linewidth]{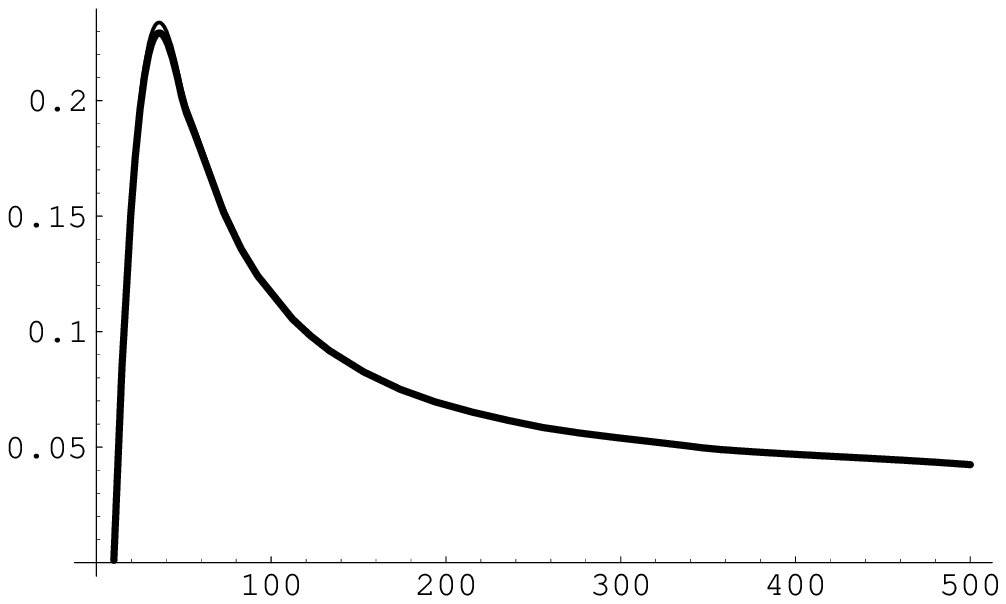}
   }
    \subfloat[]
 {
  \rotatebox{90}{\hspace{-0.0cm} {$t_{coh} \longrightarrow$}}
  \includegraphics[clip,width=0.4\linewidth]{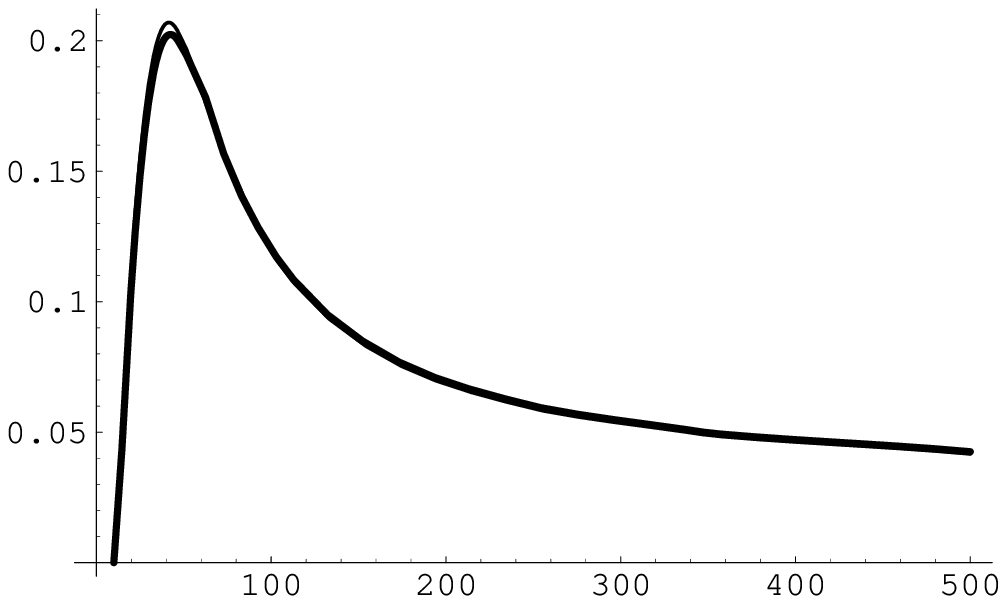}
 }\\
 \hspace{-0.0cm} {$m_{\chi} \longrightarrow$ GeV}
 \caption{The same as in Fig. \ref{fig:Ftotalt} in the case of the target $^{127}$I. Now at the bottom
 the results shown correspond to $Q_{\text{thr}}=10$ keV. The dependence on the asymmetry parameter $\beta$ is not
 visible.
}
 \label{fig:Itotalt}
 \end{center} 
  \end{figure}

\begin{figure}[tbp]
 \begin{center}
    \subfloat[]
 {
 \rotatebox{90}{\hspace{-0.0cm} {$h_{coh} \longrightarrow$}}
 \includegraphics[clip,width=0.4\linewidth]{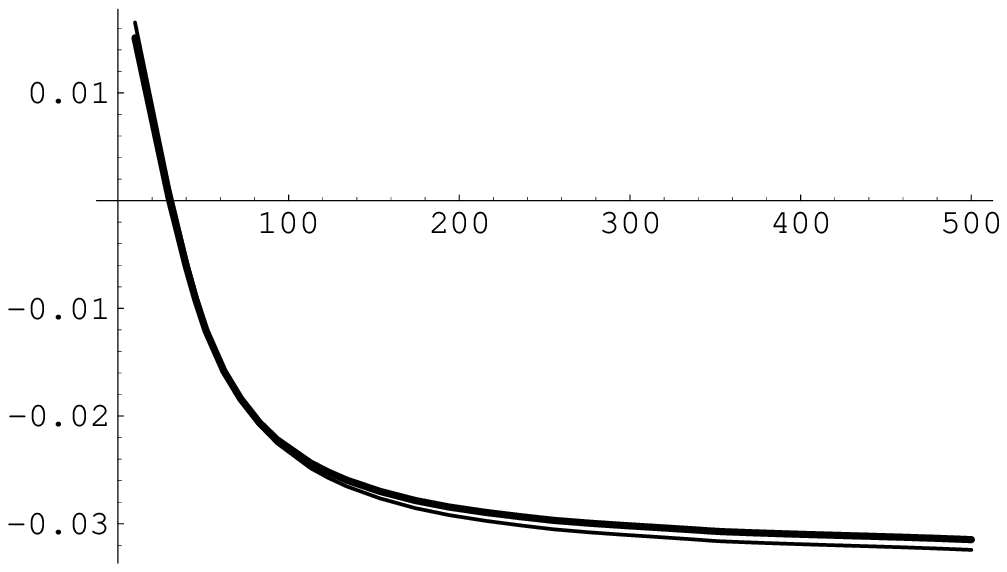}
 }
  \subfloat[]
 {
  \rotatebox{90}{\hspace{-0.0cm} {$h_{coh} \longrightarrow$}}
  \includegraphics[clip,width=0.4\linewidth]{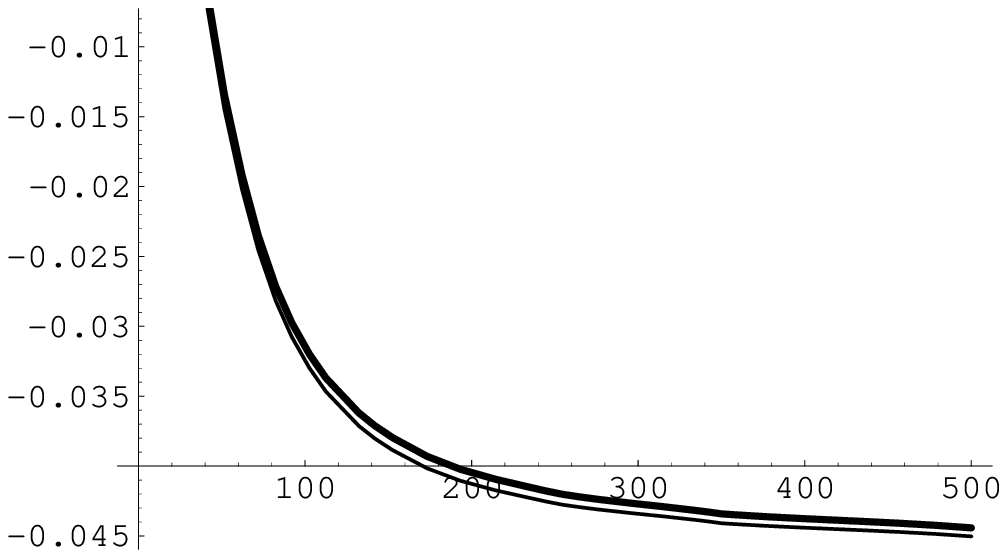}
}\\
    \hspace{-0.0cm} {$m_{\chi} \longrightarrow$ GeV}\\
     \subfloat[]
 {
  \rotatebox{90}{\hspace{-0.0cm} {$h_{coh} \longrightarrow$}}
   \includegraphics[clip,width=0.4\linewidth]{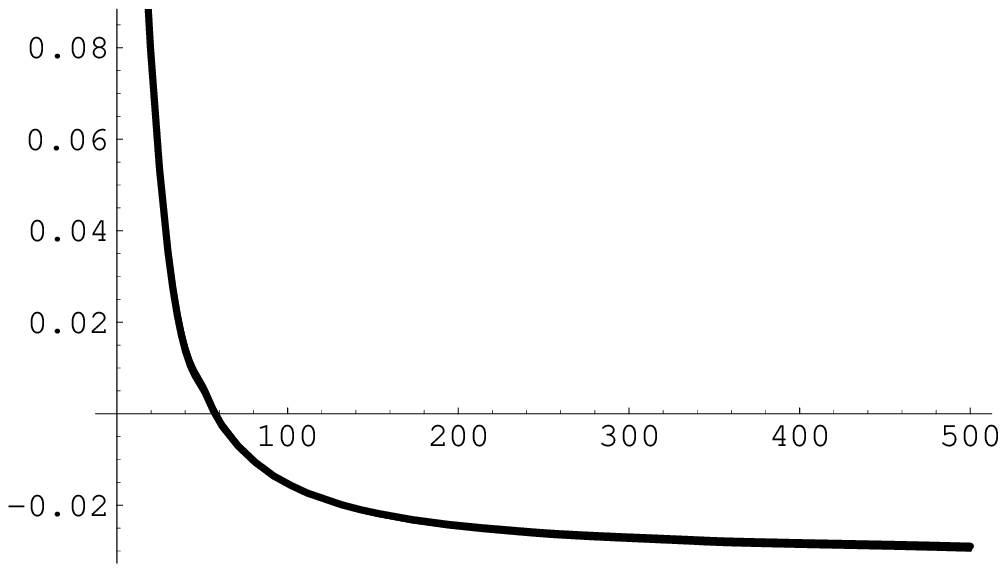}
   }
    \subfloat[]
 {
  \rotatebox{90}{\hspace{-0.0cm} {$h_{coh} \longrightarrow$}}
  \includegraphics[clip,width=0.4\linewidth]{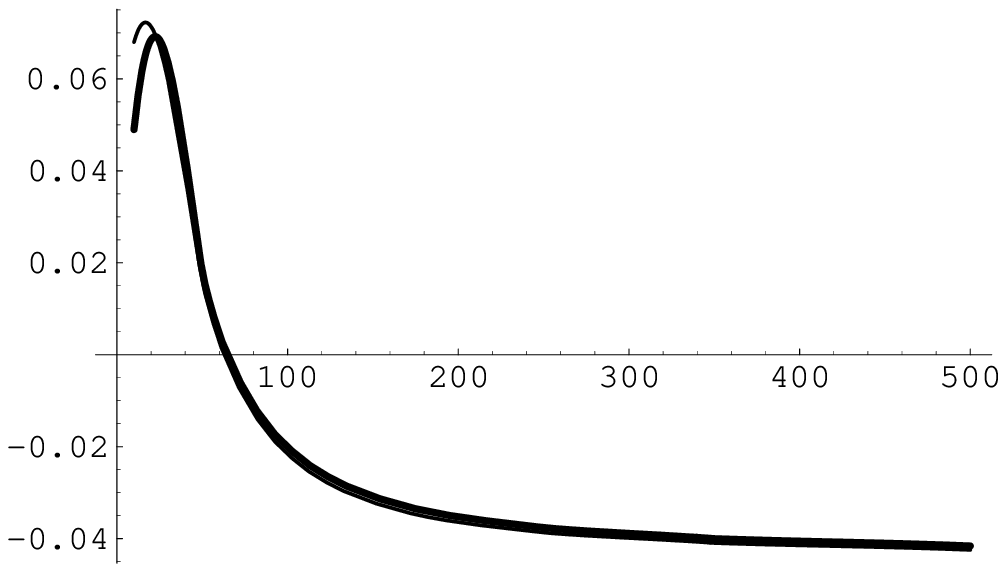}
 }\\
 \hspace{-0.0cm} {$m_{\chi} \longrightarrow$ GeV}
 \caption{The same as in Fig. \ref{fig:Itotalt} in the case of $h_{\text{coh}}$. The dependence on the asymmetry parameter $\beta$ is barely visible.
}
 \label{fig:Itotalh}
 \end{center} 
  \end{figure}

Even though the actual results have been obtained for  $^{127}$I, the 
situation is similar for other intermediate targets. The nuclear form factor employed was obtained in the shell model description of the target and is shown in
 Fig. \ref{sqformf127}.
  In the case of the target $^{127}$I the obtained results  are shown
  in Figs.~\ref{fig:Itotalt}  and \ref{fig:Itotalh} for $Q_{\text{min}}=0$ at the top and $Q_{\text{min}}=10$ keV at the bottom.
  \section{Discussion}
In our formalism the quantities which are affected by the velocity
distribution, are the parameters $t$, describing the total
time-averaged rate, and $h$ describing the amplitude of the time
dependent rate (modulation effect). These parameters also depend on
the target nucleus as well as the WIMP mass. In this work they have
been obtained in the case of the coherent scattering, but they are not
expected to be very different in the case of the spin mode. Our
results can be summarized as follows:
\begin{enumerate}
\item The time average rate.\\ 
The
parameter $t_{coh}$ in the case of symmetric velocity distribution ($\beta =0$)
is not very different from that obtained in various other models, like
those obtained in the Eddington approach
\cite{EDDIN,UK01,VEROW06,Belli:2002yt}.  The obtained results depend
on the choice of the target. Thus:
\begin{itemize}
\item The parameter $t$ for a light target\\ It is at first a
decreasing function of the WIMP mass and eventually becomes a constant
for heavy WIMPs, which for the realistic VDF
is approximately 20 percent lower than for the M-B case. This limiting
value is not much affected by threshold effects.
The value of the parameter $t$ at low WIMP masses is, as expected,
sensitive to the attained energy threshold. The dependence on the
asymmetry parameter is small (few percent) in both distributions.

\item The parameter $t$ for a heavy target.\\ The general behavior is
similar to the above except that now the limiting value is an order of
magnitude lower. This may partly offset the advantages offered by the
favorable nuclear mass dependence of the event rate, which is
proportional to $\sim \mu_r^2 A\approx A^3$ (see
Eqs.~\eqref{fullrate1}-\eqref{fullrate2}), not included in the
plots. A threshold of 10 keV further reduces the rate by about a
factor of two.  At low WIMP mass the advantage of employing an
intermediate target is expected, since $t$ is only about a factor of
three down from that of a light target.  The effect of the realistic
VDF is small (from few to 10 percent).  The effect of asymmetry is
close to unobservable.
\end{itemize}
\item The modulation effect.\\
We can now focus on two issues:
\begin{itemize}
\item The modulation of the differential rate.\\ All the previous
 velocity distributions imply a modulation amplitude which i)
 increases with energy transfer, and ii) changes sign as we go from
 low to high energy transfers. Also, iii) the slope and the point of
 the change of sign depend on the WIMP mass and the nature of the
 target.\\ In the case of the present realistic velocity all the above
 hold true. The effect of the velocity asymmetry is quite small.
\item The total modulation amplitude:\\ i) In the case of a light
target both distributions predict a similar modulation amplitude. Its
sign is positive, i.e.~the maximum is around June 3rd. The effect of
asymmetry is small, but, when comparing the two models, it tends to go
in opposite directions. Unfortunately the predicted difference between
the maximum and the minimum at zero threshold cannot exceed $4\%$. It
may double for an energy threshold of about 5 keV.

ii) In the case of a heavy-intermediate target the modulation becomes
negative for both models as the WIMP mass increases. Thus the event
rate for both models may attain a maximum in December.  The difference
between the maximum and the minimum is now quite a bit higher, i.e. it
can be as high as $9\%$ depending on the WIMP mass. For a light WIMP
mass the modulation is quite small at zero threshold. In the presence
of energy thresholds it tends to increase.  This is understandable,
since the modulation here is the ratio of two quantities (see
Eq.~\ref{Eq:h})) and the denominator, which determines the average
event rate, falls much faster then the numerator in the presence of a
threshold. The modulation can be quite sizable, especially for light
WIMPs. With a 10 keV threshold we predict differences between maximum
and minimum of about 5-15$\%$ in the case of both an M-B distribution
and the present distribution. The event rate may attain a maximum in
June for light WIMP's.  The effect of the velocity asymmetry is virtually
negligible.
 \end{itemize}
\end{enumerate} 
\section{Conclusions}
The event rate for direct WIMP detection depends on the nucleon cross
section (which in turn depends on the assumed particle model), the
WIMP mass, the structure of the target, and the WIMP velocity
distribution. In the present study we focused on the last aspect by
considering a realistic velocity distribution similar to the ones
extracted from numerical $N$-body simulations, here parametrized by
the Tsallis shape. Generally speaking the calculations using the
M-B distribution and the present distribution yield similar
results. This is particularly true of the time independent part of the
event rate. Some differences show up, however, in the prediction of
the time-dependent event rate, described by the modulation amplitude
$h$ (the difference between the maximum and the minimum event rates
being $2|h|$).  For intermediate targets at zero threshold the
modulation amplitude obtained in the present calculation is small for
light WIMPs, $|h|<2 \% $, but it can reach values of $|h|\approx5 \%$
for heavy WIMPs. For light WIMPs in the presence of an energy
threshold of a few keV the obtained values of $|h|$ may become larger,
but this occurs at the expense of the number of counts. The predicted
effect on the event rates of the asymmetry parameter of the velocity
distribution is relatively small.

It will be interesting to explore the possible consequences of the
realistic velocity distribution considered in this work on the event
rates of 
experiments which measure the direction of the recoiling nucleus
\cite{DRIFT,VERFAE06,spoo07}.

\section{acknowledgments}
SHH would like to thank Nicolao Fornengo and Stefano Scopel for early encouraging discussions. JDV would
like to thank the European Network of Theoretical Astroparticle Physics
ENTApP ILIAS/N6 under contract number RII3-CT-2004-506222 for financial
support. The Dark Cosmology Centre is funded by the Danish National Research Foundation.

\end{document}